\documentclass[twocolumn,showpacs,preprintnumbers,amsmath,amssymb]{revtex4-1}
\usepackage{epsfig}
\usepackage{epsfig,amsmath}
\usepackage{graphicx}
\usepackage{dcolumn}

\usepackage{stmaryrd}
\usepackage{mathrsfs}
\usepackage{pifont}
\usepackage{amsthm}
\usepackage{amssymb}
\usepackage{amsmath}
\usepackage{epsfig,amsmath}
\usepackage{graphicx}
\usepackage{dcolumn}
\usepackage{bm}

\newcommand{\beq}{\begin{equation}}
\newcommand{\eeq}{\end{equation}}
\newcommand{\beqa}{\begin{eqnarray}}
\newcommand{\eeqa}{\end{eqnarray}}

\newcommand{\ga}{\gamma} 
\newcommand{\da}{\dagger} 
\newcommand{\al}{\alpha} 
 \newcommand{\om}{\omega}
\newcommand{\de}{\delta} 
\newcommand{\la}{\langle}
\newcommand{\ra}{\rangle}
\newcommand{\non}{\nonumber}

\def\jpb#1{{ J.\ Phys.\ B} {\bf#1}}

\def\pra#1{{ Phys.\ Rev. A\/} {\bf#1}}
\def\prb#1{{ Phys.\ Rev. B\/} {\bf#1}}

\def\prl#1{{ Phys.\ Rev.\ Lett.} {\bf#1}}
\def\sci#1{{ Science} {\bf#1}}

\def\nat#1{{ Nature} {\bf#1}}

\begin{document}

\title{ Uhrig Dynamical Control of a Three-Level System Via Non-Markovian Quantum State Diffusion}

\author{Wenchong Shu$^1$\footnote{Email address: wshu@stevens.edu}, Xinyu Zhao$^1$,
Jun Jing$^{1,2}$, Lian-Ao Wu$^3$ and Ting Yu$^1$\footnote{Email address: ting.yu@stevens.edu}}

\affiliation{$^1$Center for Controlled Quantum Systems  and  Department of Physics and Engineering Physics,
Stevens Institute of Technology, Hoboken, New Jersey  07030,  USA\\
$^2$Institute of Theoretical Physics and Department of Physics,
Shanghai University, Shanghai 200444, China\\
$^3$IKERBASQUE, Basque Foundation for Science, 48011 Bilbao, Spain,\\
and Department of Theoretical Physics and History of Science, Basque Country University (EHU/UPV), Post Office Box 644, 48080 Bilbao, Spain
 }

\date{\today}

\begin{abstract}
In this paper, we use the quantum state diffusion (QSD)
equation to implement the Uhrig dynamical decoupling (UDD)
 to a three-level quantum system coupled to a non-Markovian
 reservoir comprising of infinite numbers of
degrees of freedom. For this purpose, we first reformulate the non-Markovian
 QSD to incorporate the effect of the external control fields. With this
 stochastic QSD approach,  we demonstrate that an unknown state of
the three-level quantum system can be universally protected against
both colored phase and amplitude noises when the control-pulse
sequences and control operators are properly designed.
The advantage of using non-Markovian quantum state diffusion
equations is that the control dynamics of open quantum systems
can be treated exactly without using Trotter product formula and
be efficiently simulated
even when the environment comprise of infinite numbers of
degrees of freedom. We also show how the control
efficacy depends on the environment memory time and the designed time points
of applied control pulses.


\end{abstract}


\maketitle

\section{Introduction}

Few-level atomic and molecular systems play crucial roles in quantum control and quantum information processing.
For example, qubits have great advantages in certain computational tasks compared to classical bits because of quantum coherence
and quantum entanglement. However,
for an open quantum system, mutual effects due to the coupling between the system and environment are inevitable and result in very complex
reduced dynamics including dissipation, fluctuation, decoherence and disentanglement \cite{decoherence1,decoherence1a,Entanglement1,YuEberly1,decoherence1b,decoherence2}.

Inspired from the Hahn spin echo in nuclear magnetic resonance (NMR) \cite{Hahn}, dynamical decoupling (DD) was
shaped into a useful tool to mitigate the decoherence of a quantum system coupled to an environment \cite{CarrPurcell,MeiboomGill,ViolaLloyd1,ViolaLloyd2,WuMark,KhodjastchLidar1,KhodjastchLidar2}.
Recently, different aspects of DD have been studied with significant progress; It has been shown that
by using aperiodic control pulses, the so-called Uhrig dynamical decoupling (UDD) scheme, a single qubit
in the pure dephasing spin-boson
model can maintain its coherence to the $N^{th}$ order by using only $N$ or $N+1$ pulses \cite{Uhrig1,Uhrig2,YangLiu1}. Furthermore,
Ref. \cite{WestLidar1} constructed two layers of nesting UDD that can protect a single qubit against both dephasing
and relaxation. In addition, the multi-layer nesting UDD and continuous DD schemes are designed to not only preserve quantum coherence and
 the entanglement of two-qubit systems \cite{GordonKurizki, MukGong2, KaoMou, PanGong, ChaudhryGong},  but also to protect multi-qubit systems \cite{WangLiu} and
quantum gates \cite{WestGyure}.  Studies on higher-order effects have also been done for non-ideal pulses
\cite{PasiniUhrig2, PasiniUhrig3}, as well as optimized pulses under different environment
noises \cite{GordonLidar, ChenLiu, PasiniUhrig4}. In experiments, the remarkable performance of UDD in prolonging the life time of a
single-qubit state in various models has been studied \cite{BiercukBollinger, DuLiu, LangeHanson, AjoySuter}.

The purpose of this paper is to develop a non-Markovian QSD approach that can naturally incorporate the external
control fields in the framework of Uhrig's dynamical coupling approach. We present a UDD scheme with new control
operators that may be used to control an unknown
state of the three-level system (spin-1 system) against dissipation and pure phase decoherence.
Our strategy is first to derive a time-local non-Markovian quantum-state-diffusion (QSD) equation for a single three-level dephasing model
and dissipative model under the external DD control fields, then we show that  the non-Markovian QSD
 can be systematically solved to simulate Uhrig control dynamics of a three-level system under the influence of
 dephasing and dissipative noises. One of important features of our stochastic approach is that we don't make
 any assumptions on the system-environment coupling and the size of environment. In this way, we can consistently
 solve the control dynamics without invoking the Trotter product formalism.

If quantum systems are coupled to small environments,  it is straightforward to use the direct numerical simulations to solve the combined system
and environment such that the UDD scheme can be efficiently implemented \cite{MukGong2}.  When the open system is coupled
 to a large environment consisting of finite or infinite numbers of degrees of freedom, one has to invoke an efficient quantum approach
 to deal with the open system dynamics allowing dealing with both non-Markovian and Markov environments.
 Since the control pulses applied to the open quantum system are typically implemented in
a very short time scale and high intensity that is much stronger than the system-environment coupling, it is usually assumed
there is no coupling between the system and the environment when the control pulses are applied. Therefore in
 the previous work, the dynamics of the controlled
open system behaviors like the quantum jump process \cite{DalibardMolmer,Uhrig1,YangLiu1}. In this paper, without the above
assumption, we treat the total system plus environment with control as a integrated and consistent entirety to solve its exact dynamics by using the non-Markovian QSD
equation initially proposed in \cite{StrunzGisin}.  Derived directly from an underlying microscopic model
irrespective of environment memory time and coupling strength, the stochastic QSD equation is a useful approach for
solving several models exactly \cite{YuStrunz2,Model,Yu2004,YuStrunz3, JJing1}. As shown below, the QSD equation can
provide a systematic tool to deal with the non-Markovian quantum open system under the UDD control fields. The advantage of using
non-Marokivian QSD is its numerical power and its versatility in dealing with varied environmental sizes ranging from a few degrees
of freedom to infinite numbers of degrees of freedom in an arbitrary non-Markovian regimes.

The organization of this paper is as follows: In Sec.~II, we present the physical models and our control strategy for a three-level atomic system
interacting with a phase noise environment. We introduce a modified non-Markovian quantum state diffusion equation (QSD)
without time-dependent coupling operator and apply it to the dynamical control of the three-level system.
In Sec.~III, we present the nesting sequences to control the three-level system coupled to the dissipative environment.
We derive a set of dynamic equations for the coefficients of the quantum state diffusion equation. 
Lastly, Sec.~IV makes discussions and concludes the paper.

\section{Modified QSD equation for dephasing noise}

In this section, we study the DD scheme for an open quantum system involving a three-level atom coupled linearly to a general bosonic environment
consisting of a set of bosonic operators $b_n, b_n^\da$ satisfying $[b_n, b_m^\dag]=\delta_{nm}$. The total Hamiltonian may be written as (setting
$\hbar\equiv1$):
\beq H_{\rm tot}=H_{\rm sys}+H_{\rm env}+H_{\rm int}, \eeq
with the three terms:
\begin{align}
H_{\rm sys}  &  =\omega J_{z},\nonumber\\
H_{\rm env}  &  =\sum_{n}\omega_{n}b_{n}^{\dagger}b_{n},\nonumber\\
H_{\rm int}  &  =\sum_{n}J_{z}(g_{n}b_{n}^{\dagger}+g_{n}^*b_{n}),
\end{align}
where the system has three energy levels $E_0=-\om$, $E_1=0$ and $E_2=\om$, $J_z=|2\ra\la2|-|0\ra\la 0|$ is the coupling
operator and $g_n$ are the coupling constants between the three-level atom and the environmental modes. This is a pure
dephasing model where there are no population transitions between the energy levels as the interaction Hamiltonian
commutes with the system Hamiltonian.

For this pure dephasing noise,  it is not difficult to show that the iterative pulses of Uhrig's type can preserve  an arbitrary initial state.
More precisely, for the DD scheme, the control sequence consisting of $N$ instantaneous pulses over a  duration
of time $T$ can be described by the following control Hamiltonian:
\beq H_{\rm ctr}(t)=\sum_{j=1}^N\frac{\pi}{2}\de(t-T_j)P, \eeq
where we choose the Uhrig DD(UDD) time intervals \cite{Uhrig1, Uhrig2}:
\beq T_j=T\sin^2\left(\frac{j\pi}{2N+2}\right), j=1,2,\cdots,N.\eeq
The control operator $P$ for a single three-level pure dephasing model may be determined by using these two
criteria \cite{MukGong2}: $P^2=I$ and $\{J_z,P\}=0$. Therefore,  it is easy to check that the following two control operators satisfy the
required conditions:
\beq \label{Pcontrol} P=\left[\begin{array}{ccc} 0&0&1 \\ 0&1&0 \\ 1&0&0 \end{array} \right]. \eeq

In the rotating reference frame with respect to $H_{\rm ctr}$, we can obtain an effective Hamiltonian in the interaction picture.
Using the commutation relation $[P,[P,J_z]]=4J_z$,  we have
\beqa
&&  \exp[-i\int_{0}^{t}dsH_{\rm ctr}(s)](J_{z})\exp[i\int_{0}^{t}dsH_{\rm ctr}(s)]\nonumber\\
&&  =\exp[-i\frac{\pi}{2}%
{\textrm{Step}}(t)P](J_{z})\exp[i\frac{\pi}{2}
{\textrm{Step}}(t)P]\nonumber\\
&&  =p(t)J_{z},
\eeqa
where ${\textrm{Step}}(t)=j$ when $t\in[T_{j},T_{j+1})$; And
$p(t)=\pm 1$, which changes sign at the time points  $T_j$.  Now, the total Hamiltonian describing the control pulse
 sequence plus the system and environment in such a "toggling frame" \cite{ViolaKnill, UhrigLidar, NgPreskill} can be written as:
\begin{align} \label{effective H depashing under control}
\widetilde{H} &= \exp\left[-i\int_{0}^{t}H_{\rm ctr}(s)ds\right] H_{\rm tot} \exp\left[i\int_{0}^{t}H_{\rm ctr}(s)ds\right] \non\\
&= \widetilde{H}_{\rm sys}+H_{\rm env}+\widetilde{H}_{\rm int},
\end{align}
where the effective system Hamiltonian and system-environment interaction Hamiltonian are given by:
\beqa \label{flip over}
\widetilde{H}_{\rm sys}&=&\om p(t)J_{z}, \non\\
\widetilde{H}_{\rm int}&=&\sum_{n}p(t)J_{z}(g_{n}b_{n}^{\dagger}+g_{n}^*b_{n}).
\eeqa

The exact dynamics for the three-level atomic system under both the environmental noise and control pulses can be compactly described
by the non-Markovian QSD equation derived from the above total Hamiltonian in the toggling frame:
\begin{equation} \label{QSD}
\frac{\partial}{\partial t}\psi_{t}=-i\widetilde{H}_{\rm sys}\psi_{t}+Lz_{t}^{\ast}\psi
_{t}-L^{\dagger}
{\displaystyle\int\nolimits_{0}^{t}}
ds\alpha(t-s)\frac{\delta\psi_{t}}{\delta z_{s}^{\ast}},
\end{equation}
where $L=p(t)J_{z}$ is the modified system Lindblad operator incorporating both the effects of environment and the external control pulses.
Note that the correlation function  $\alpha(t-s)=\sum_n |g_n|^2e^{-i\om_n(t-s)}$ is arbitrary and $z_t^*=-i\sum_n g_nz_n^*e^{iw_n t}$
is a complex Gaussian process satisfying $M[z_t]=M[z_tz_s]=0$ and $M[z_tz_s^*]=\al(t,s)$. Here $M[\, .\, ]$ denotes the statistical average
 over the classical Gaussuan noise $z_t$. When $\al(t,s)=\Gamma \delta(t-s)$, the noise $z^*_t$ reduces to the memoryless white noise.
 It should be noted that the above non-arkovian QSD equation represents a new type of QSD with a time-dependent Lindblad operator.

In order to solve the non-Markovian QSD equation, we may rewrite the functional derivative term as $\frac{\delta}{\delta z_{s}^{\ast}}\psi_{t}(z^*)=O(t,s,z^*)\psi_{t}(z^*)$,
where $O(t,s,z^*)$ is a time dependent operator acting on the system Hilbert space. The equation of motion for   $O(t,s,z^*)$ can be obtained by
by using the consistency condition \cite{DiosiStrunz}:
\begin{equation} \label{consistency condition}
\frac{\partial}{\partial t}O=[-i\widetilde{H}_{sys}+Lz_{t}^{\ast}-L^{\dagger}\overline
{O},O]-L^{\dagger}\frac{\delta}{\delta z_{s}^{\ast}}\overline{O},
\end{equation}
where $\overline{O}(t,z^*)=\int_0^tds\al(t,s)O(t,s,z^*)$.

For the three-level dephasing model under DD control described by Eq.~(\ref{effective H depashing under control}), it can be shown that the exact $O$ operator
is simply given by:
\beq
O(t,s,z^*)=f(t,s)J_z,
\eeq
with $f(t,s)=p(s)$. Therefore, the explicit non-Markovian QSD equation can be compactly written into:
\beq \label{QSD for controlled depha}
\frac{\partial}{\partial t}\psi_{t}(z)=(-i\om+z_t^*+F(t)J_z)p(t)J_z\psi_{t}(z),
\eeq
where $F(t)=\int_0^tds\al(t,s)p(s)$. By calculating the statistical average over many realizations of trajectory generated by the stochastic process $z^*_t$,
one can recover the density operator of the three-level system:
\beqa \rho_t &=&M[|\psi_t\ra\la\psi_t|] \non\\
&=& \int\frac{dz^2}{\pi}e^{-|z|^2}|\psi_t(z^*)\ra\la\psi_t(z^*)|.
\eeqa

It is known that the exact QSD equation can be applicable to an open system model with an arbitrary correlation function $\al(t, s)$.
In order to investigate how the environment memory time affects the effectiveness of the DD control, in the following numerical simulations,
we model the environmental noise as the Ornstein-Uhlenbeck process with the correlation function although our approach is valid for an arbitrary
correlation function:
\beq
\alpha(t,s)=\frac{\gamma}{2}
e^{-\gamma\left\vert t-s\right\vert }.
\eeq
where $\gamma$ represents essentially the environmental bandwidth, hence the environment memory time scale can be represented
by the parameter $1/ \ga$.  The advantage of choosing the Ornstein-Uhlenbeck process is that the Markov limit is simply dictated by
the single parameter $\gamma$. When $\ga \rightarrow \infty$, $\alpha(t,s) \rightarrow \de(t-s)$ recovering the Markov limit.
Typically, a finite (small)  $\ga$ represents a non-Markovian regime. Note that,  for the Ornstein-Uhlenbeck noise, the equation of motion
for $F(t)$ is simply given by:
\begin{equation}
\label{Ffunction}
\frac{d}{dt}F(t)=\frac{\gamma}{2}p(t)-\gamma F(t).
\end{equation}
With (\ref{Ffunction}),  the QSD Eq.~(\ref{QSD for controlled depha}) is fully determined. We first compute the fidelity and the angular momentum time evolution
with different numbers of control pulses. The plots are shown in Fig.~1.

\begin{figure}[htb]
\centering\includegraphics[width = 7cm]{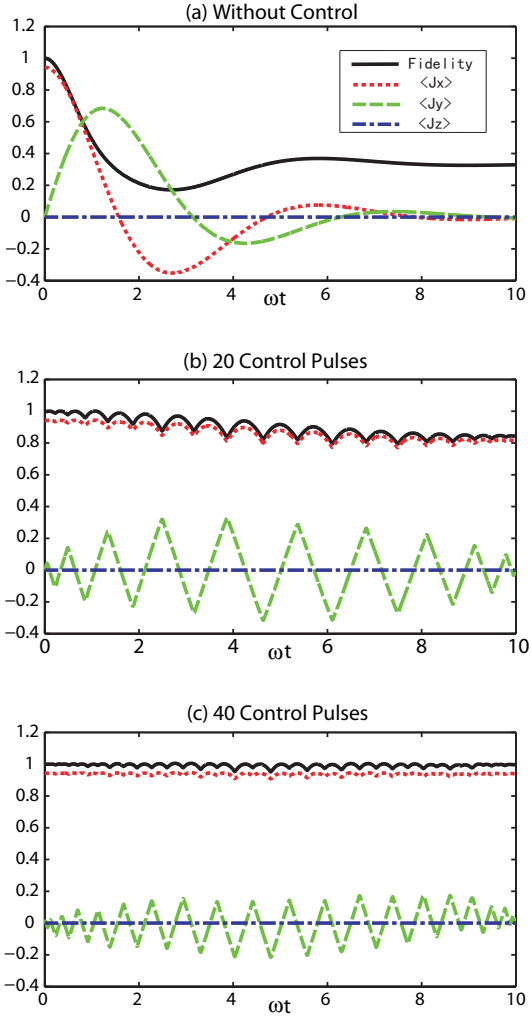}
\caption{Ensemble average of Fidelity and $\la \vec{J}\ra$ over 2000 trajectories of the single qutrit dephasing model under different UDD control sequences. We choose the initial state $|\psi_0\ra=\frac{1}{\sqrt{3}}(|0\ra+|1\ra+|2\ra)$ and the environment memory parameter $\ga=1$. The black solid line represents the Fidelity, red dotted line denotes $\la J_x \ra$, green dashed line denotes $\la J_y \ra$ and blue dot-dashed line denotes $\la J_z\ra$. }
\label{dephasing_0,20,40pulses}
\end{figure}

Without the UDD control, as shown in Fig.~\ref{dephasing_0,20,40pulses}(a),  the three-level system under the pure dephaing relaxation will evolve into a complete mixed final state.  Here we choose the environment memory parameter $\ga=1$ which stands for a moderate non-Markovian regime
\cite{JJing1}.  In Fig.~\ref{dephasing_0,20,40pulses}(a), the dynamics of $\la J_x \ra$ and $\la J_y \ra$ is shown to exhibit a few oscillations before
reaching their final values.

However, with the UDD control sequence applied to this three-level system, it is shown that the system's fidelity for the given initial state
is well protected as illustrated in \ref{dephasing_0,20,40pulses}(b) and (c). Clearly, a better control result can be achieved
 if more controlled pulses are used in the control scheme.

The control processes may be better understood by inspecting the shapes of $\la J_x \ra$ and $\la J_y \ra$ curves.  At each UDD time point, the
effect of the control pulse is simply to  change the sign of operator acting on the system given by Eq.~(\ref{flip over}). As result,  after each single control
pulse, the the mean angular momentum is modified towards its opposite direction. On average both $\la J_y \ra$ and $\la J_x \ra$ are effectively preserved
in the presence of the noise as illustrated by the temporal evolution of the three-level quantum system.

\begin{figure}[htb]
\centering\includegraphics[width = 8cm]{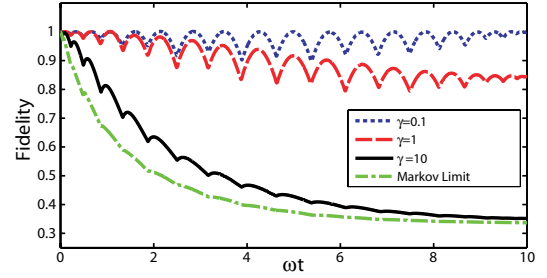}
\caption{Fidelity evolution of the single qutrit dephasing model under 20 control pulses UDD sequence with different environment memory parameter $\ga$. The initial state $|\psi_0\ra=\frac{1}{\sqrt{3}}(|0\ra+|1\ra+|2\ra)$.}
\label{dephasing_different_gamma}
\end{figure}

It is interesting to know how the effectiveness of the UDD control is affected by the environment memory times \cite{EscherKurizki}. The results are shown in Fig.~\ref{dephasing_different_gamma}. Clearly, the transition of dynamics from non-Markovian to Markov regimes is dictated by environment memory time $\tau=1/ \ga$.  In the case of small $\ga$, which stands for a long environment memory time, the more pronounced effects of control pulses on the three-level system is expected. In fact, as seen from the example,  a longer coherence
time may be preserved when the environment has a long memory time.  Consequently, the fidelity will be efficiently protected. In contrast,
when the system approaches Markov limit as the memory time  $\tau$ becomes shorter and shorter ({\t i.e.,}  $\ga \gg 1$),  ineffectiveness of the control pulses can be
easily observed. This is easy to understand as the rapid coherence decay has rendered the system effectively a classical ensemble before the external control pulses
take effect (see black solid line and green dash-doted line in Fig.~\ref{dephasing_different_gamma}). This has clearly demonstrated
that an engineered environment with a longer memory time typically conduces to a better control of  the system when the UDD pulses are applied \cite{LiuGuo}.

\begin{figure}[htb]
\centering\includegraphics[width = 8cm]{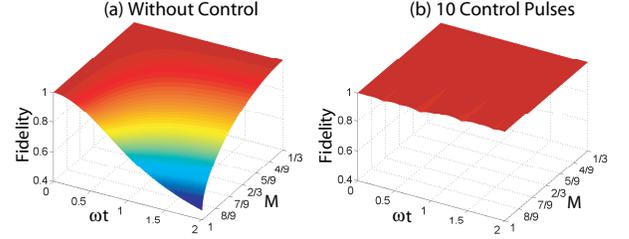}
\caption{ Fidelity time evolution of a single-qutrit dephasing model with the initial state $\rho_0=\frac{1-M}{2}I_3+\frac{3M-1}{2}|\psi_0\ra\la\psi_0|$, where $M$ is the degree of mixing and $|\psi(0)\ra=\frac{1}{\sqrt{3}}(|0\ra+|1\ra+|2\ra)$. The environment memory time $t=1/\gamma$ with $\ga=1$.}
\label{dephasing_mixed}
\end{figure}

Another interesting situation is that the initial state of the three-level system is a mixed state. Since any initial mixed state can be represented as $\rho_0=\sum_\lambda c_\lambda |\psi_{\lambda,0}\ra\la\psi_{\lambda,0}|$, then the density matrix at time $t$ is simply  expressed in terms of QSD solutions: $\rho_t=\sum_\lambda c_\lambda M[ |\psi_{\lambda,t}\ra\la\psi_{\lambda,t}|]$, where $|\psi_{\lambda,t}\ra$ is governed by the QSD Eq. (\ref{QSD for controlled depha}).

As an illustration, we consider the initial state as a Werner-like state \cite{Werner, ShuYu1}:
\beq \label{Werner-like state}
\rho_0=\frac{1-M}{2}I_3+\frac{3M-1}{2}|\psi_0\ra\la\psi_0|, \eeq
where $|\psi_0\ra= \frac{1}{\sqrt{3}}(|0\ra+|1\ra+|2\ra)$.
The parameter $M\in[1/3,1]$ describes the "degree of mixture" of the initial qutrit state $\rho_0$,
when $M=1/3$ it is a maximally mixed state, while $M=1$ it reduces to the pure state $|\psi_0\ra$.

The time evolution of fidelity is illustrated in Fig.~\ref{dephasing_mixed}.
Left 3-D picture (a) plots fidelity dynamics without UDD control.  It is seen
that the fidelity  decays faster for an initial qutrit state with a higher degree of purity.
However, when applying only 10 UDD pulses, the fidelity over time scale considered here is well protected
as shown in Fig.~\ref{dephasing_mixed}(b).

\section{Modified QSD equation for dissipative noise}

In this section we will consider a there-level system coupled to a dissipative noise \cite{JJing1}, the total Hamiltonian of the system plus environment
can be described as:
\beq
H_{\rm tot} =\omega J_{z}+\sum_{n}\omega_{n}b_{n}^{\dagger}b_{n}+\sum_{n}(g_{n}J_{-}b_{n}^{\dagger}+g_{n}^*J_{+}b_{n}),  \label{totdissi}
\eeq
where $J_{-}=\sqrt{2}(|0\ra\la1|+|1\ra\la2|)$ and $J_{+}=\sqrt{2}(|2\ra\la1|+|1\ra\la0|)$. Clearly, without control, the initial qutrit state will lose
its coherence quickly by dissipation.  In order to suppress the decoherence and protect the fidelity of the initial state of the three-level
system,  it is shown that two layers of DD control sequences are necessary. The control Hamiltonian can be written as:
\beqa
&&H_{\rm ctr}=H_{c1}+H_{c2} \non\\
&&=\sum_{j=1}^{N_1}\frac{\pi}{2}\de(t-T_j)P+
\sum_{j=1}^{N_1}\sum_{k=1}^{N_2}\frac{\pi}{2}\de(t-T_{j,k})Q, \label{controldissi}
\eeqa
where $H_{c1}$ is the outer layer UDD sequence consisting of number $N_1$ of $P$ pulses as in Eq. (\ref{Pcontrol}) of Sec. II. And $H_{c2}$ is the inner
layer UDD sequence of $Q$ pulses which would be applied at the time points \cite{WestLidar1}:
\beq T_{j,k}=T_j+(T_{j+1}-T_j)\sin^2\left(\frac{k\pi}{2N_2+2}\right). \eeq

In the case of  dissipative noise, it can be shown that the two control operators $P$ and $Q$ are given by:
\beq
P=\left[\begin{array}{ccc} 0&0&1 \\ 0&1&0 \\ 1&0&0 \end{array} \right],
Q=\left[\begin{array}{ccc} 1&0&0 \\ 0&-1&0 \\ 0&0&1 \end{array} \right],  \eeq
where the new control operator Q can be derived from the following criteria \cite{MukGong2}: $Q^2=I$, $[P,Q]=[J_z,Q]=0$ and $\{J_-,Q\}=\{J_+,Q\}=0$.

 In order to solve the system dynamics of the three-level dissipative model under the UDD control scheme, we need to use the rotating frame transformations twice, one for
 the control Hamiltonian $H_{c1}$ and the other for the control Hamiltonian $H_{c2}$, respectively (see Appendix A for details). The final effective total Hamiltonian in the new togging frame are given by:
 \beq \label{effective H dissi under control}
  \widetilde{H}=\widetilde{H}_{\rm sys}+ \sum_{n}\omega_{n}b_{n}^{\dagger}b_{n}
 + \sum_{n}(g_{n}b_{n}^{\dagger}L+g_{n}^*b_{n}L^\da), \eeq
 where
 $\widetilde{H}_{\rm sys}=p(t)\om J_z$ and the effective Lindblad operator is:
 \beq
L=l_1(t)J_-+l_2(t)J_+,
\eeq
where the coefficients are:
\beqa
l_1(t)&=&q(t)\frac{1+p(t)}{2}, \non\\
l_2(t)&=&q(t)\frac{1-p(t)}{2},
\eeqa
Here the piecewise functions $p(t)=\pm1$ (the values change at time points $T_{j}$) and $q(t)=\pm1$ (the values change at time points $T_{j,k}$). Again, we see that
the effect of the external control field is represented by the time-dependent Lindblad operator.

\begin{figure}[htb]
\centering\includegraphics[width = 8cm]{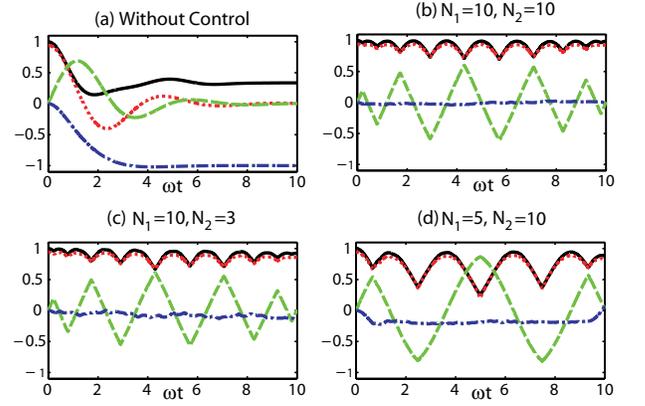}
\caption{ Ensemble average of Fidelity and $\la \vec{J}\ra$ over 2000 trajectories of the single qutrit dissipative model under different UDD control sequences. We choose the initial state $|\psi_0\ra=\frac{1}{\sqrt{3}}(|0\ra+|1\ra+|2\ra)$ and the environment memory parameter $\ga=1$. The black solid line represents the Fidelity, red dotted line denotes $\la J_x \ra$, green dashed line denotes $\la J_y \ra$ and blue dot-dashed line denotes $\la J_z\ra$. }
\label{dissi_different_pulses}
\end{figure}

For the effective Hamiltonian given in Eq.~(\ref{effective H dissi under control}), we can establish a modified QSD Eq.~(\ref{QSD}) with time-dependent Lindblad operators. Once the correlation function $\alpha(t,s)$ is given, the modified QSD equation for the three-level system under control can be solved numerically.  For simplicity, we use the perturbation $O$ operator \cite{YuStrunz1} corresponding to the weak noise approximation (for more details, see Appendix B),
\beqa \label{O for dissi}
O(t,s)&=&f_1(t,s)J_-+f_2(t,s)J_+ \non\\
&& + f_3(t,s)J_zJ_-+f_4(t,s)J_zJ_+, \\
\overline{O}(t,s)&=& \int^t_0ds\al (t,s)O(t,s)\non\\
&=& F_1(t,s)J_-+F_2(t,s)J_+ \non\\
&&+ F_3(t,s)J_zJ_-+F_4(t,s)J_zJ_+, \label{Obar for dissi}
\eeqa
where $F_i(i=1,2,3,4)=\int^t_0ds\al(t,s)f_i$.  It is noted that the QSD equation derived here is valid for an arbitrary correlation function.  In our numerical simulations, we always choose the Ornstein-Ulenback type of correlation function $\al(t,s)=\frac{\ga}{2}e^{-\ga|t-s|}$. By substituting Eq.~(\ref{O for dissi}) and (\ref{Obar for dissi}) into Eq.~(\ref{consistency condition}), we can derive  a set of  differential equations for the coefficients as:
\beqa
 \frac{\partial}{\partial t}F_1 &=&\frac{\ga}{2}l_1+(ip\om-l_1F_3+l_2F_4-\ga)F_1, \non\\
 \frac{\partial}{\partial t}F_2 &=&\frac{\ga}{2}l_2+(-ip\om-l_1F_3+l_2F_4-\ga)F_2, \non\\
 \frac{\partial}{\partial t}F_3 &=&(ip\om-3l_1F_1+l_1F_3-l_2F_4-\ga)F_3 \non\\
 &&+(l_1F_1-l_2F_2+l_2F_4)F_1, \non\\
 \frac{\partial}{\partial t}F_4 &=& (-ip\om-3l_2F_2+l_1F_3-l_2F_4-\ga)F_4 \non\\
 &&+(l_1F_1-l_2F_2+l_1F_3)F_2.
  \eeqa
  with $F_i(0)=0$. Finally, the QSD Equation is given by:
\beqa
\frac{\partial}{\partial t}\psi_t &&= \left[-ip\om J_z+(l_1J_-+l_2J_+)z_t^* \right. \non\\
&& -l_1F_1J_+J_--l_1F_2J_+^2-l_1F_3J_+J_zJ_- \non\\
&& \left.-l_2F_2J_-J_+-l_2F_1J_-^2-l_2F_4J_-J_zJ_+\right]\psi_t.
\eeqa

\begin{figure}[htb]
\centering\includegraphics[width = 6.5cm]{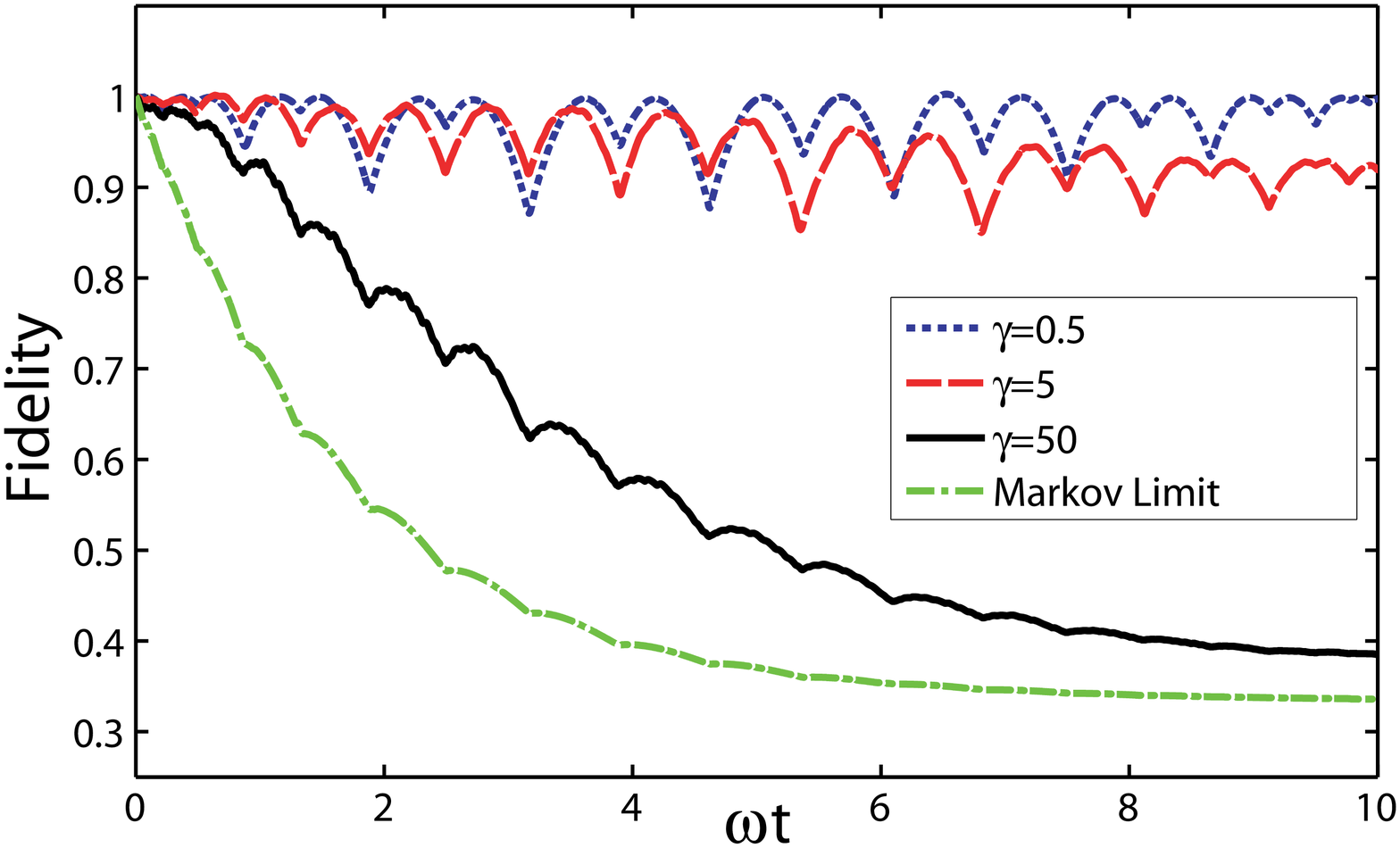}
\caption{Fidelity evolution of single qutrit dissipative model under control with different environment memory parameter $\ga$. We apply two-layer nesting UDD control sequences with the outer layer $N_1=20$ and the inner layer $N_2=10$. The initial state $|\psi(0)\ra=\frac{1}{\sqrt{3}}(|0\ra+|1\ra+|2\ra)$.}
\label{dissi_different_gamma}
\end{figure}

The numerical results with 2000 realizations are plotted in Fig.~\ref{dissi_different_pulses}. For the zero-temperature environment, without any control pulses, the spontaneous emission always causes the three-level system to decay into the ground state. So in Fig.~\ref{dissi_different_pulses} (a), when $t \rightarrow \infty$, we get $\la J_x \ra,\la J_y \ra \rightarrow 0 $ and $\la J_z \ra \rightarrow -1$. Beyond single UDD control sequence for dephasing model, by applying two layers of nesting UDD control sequences, we can also successfully resist the dissipation (Fig.~\ref{dissi_different_pulses} (a), (b) and (c)).

\begin{figure}[htb]
\centering\includegraphics[width = 8.3cm]{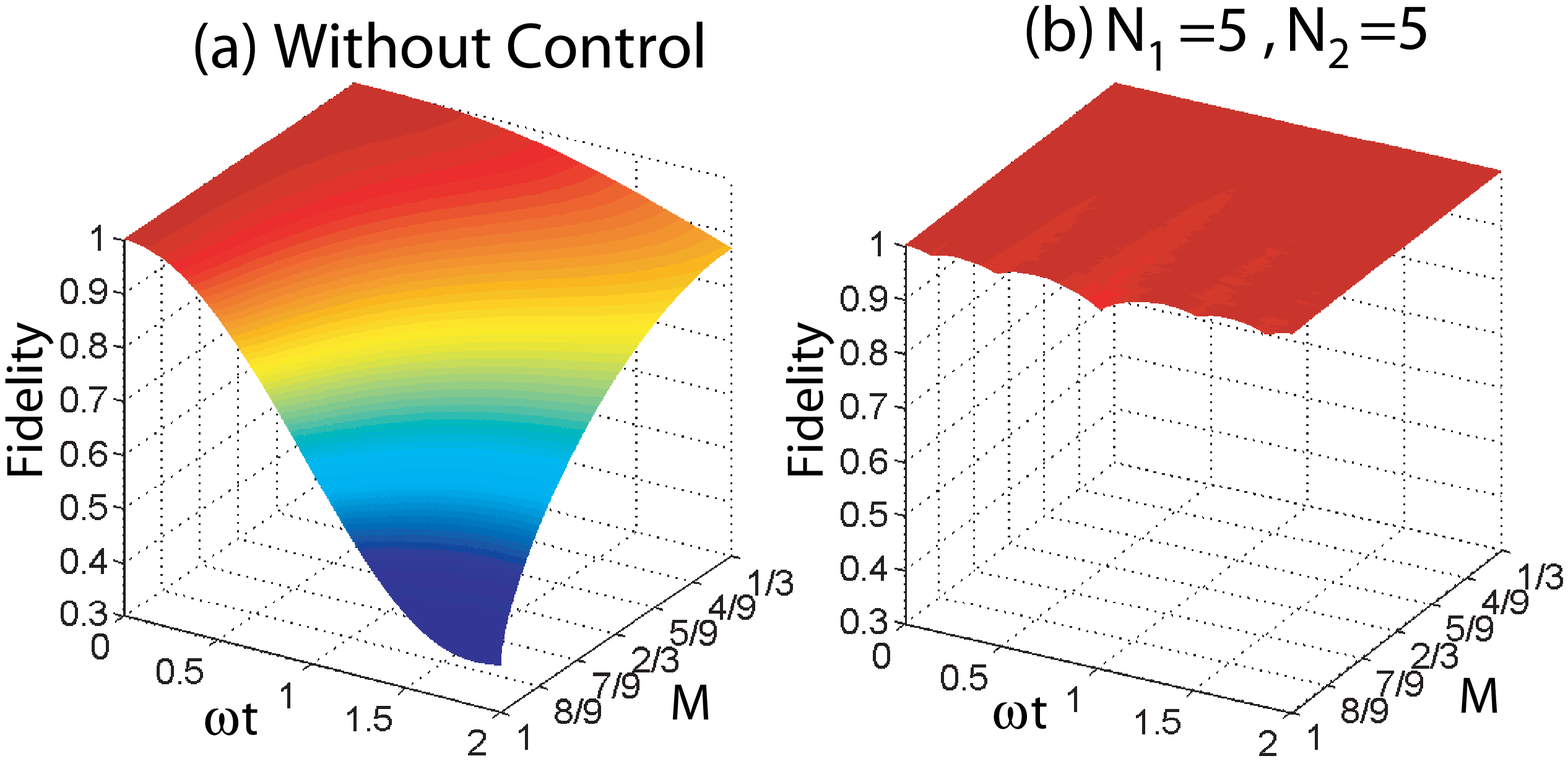}
\caption{ Fidelity time evolution of a single-qutrit dissipative model with the initial state $\rho_0=\frac{1-M}{2}I_3+\frac{3M-1}{2}|\psi_0\ra\la\psi_0|$, where $M$ is the degree of mixing and $|\psi(0)\ra=\frac{1}{\sqrt{3}}(|0\ra+|1\ra+|2\ra)$. The environment memory index $\ga=1$.}
\label{dissi_mixed}
\end{figure}

For the two-layer nesting UDD sequences, the total number of
the control pulses is $ N_{tot}=N_1+(N_1-1)N_2 $. So in the  case illustrated by Fig.~\ref{dissi_different_pulses} (b), $N_1=N_2=10$, there are totally 100 pluses applied to the dissipative model. Compared this to the dephasing model with 40 control pulses (Fig.~\ref{dephasing_0,20,40pulses} (c)), we find that the dissipative model needs more control pulses to achieve the same degree of protection. Also let us compare Fig.~\ref{dissi_different_pulses} (c) and Fig.~\ref{dissi_different_pulses} (d). There are 37 pulses in the  former case and 45 pulses in the latter case. However, the fidelity evolution and the angular momentum evolution for case (c) are much better protected than case (d). This may suggest that the number of outer layer pulses is the dominant factor of the control efficiency.

Fig.~\ref{dissi_different_gamma} shows the fidelity dynamics with different environment memory times.  Similar to the result obtained in the last section, the long memory ability will enhance the function of DD control. In Fig~\ref{dissi_different_gamma}, we use larger values of $\ga$ for the dissipative model than we did in Fig. \ref{dephasing_different_gamma}, because the total number of pulses is one order of magnitude larger than the number we used in dephasing model. This restriction causes the environment to reach the Markov limit slowly by increasing the parameter $\ga$.

Lastly, we choose the Werner-like state (Eq.~(\ref{Werner-like state})) as the initial state of the three-level system for the dissipative model. The ensemble average of fidelity time evolution with different degrees of mixture is depicted in the 3-D Fig.~\ref{dissi_mixed}. Without control, any initial state will evolve to the final steady state $|0\ra$ as time goes by, and the fidelity decays faster as the purity of the initial state increases.  Fig.~\ref{dissi_mixed}(b) demonstrates that two-layer nesting UDD sequences satisfactorily control all of the mixed states.

\section {Concluding remarks}

With both theoretical analysis and numerical simulation, we have shown that the modified quantum state diffusion approach with time-dependent Lindblad operators can be a useful tool for implementing
the UDD scheme in the situations where the open system is coupled to a large non-Markovian environment.  For a non-Markovian three-level open
system, we show that a new control strategy in DD scheme can efficiently protect an unknown three-level quantum state against both dephasing
and dissipation noises. This new set of control operators only works for ladder-type three-level quantum system with equal distance eigenvalues.
Our three-level system represents a spin-1 or angular momentum system which is of interest in many physically interesting case such as quantum cryptography and
quantum entanglement \cite{Zeilinger2001}. It is interesting to note that a more general three-level system with different energy spacings such as V-type or $\lambda$-type atoms,
a universal effective control via UUD combined with non-Markovian QSD are still possible, but it becomes much more complicated technically
and  multi-nesting sequences with different control operators have to be employed. Moreover, the exact QSD will be difficult to find, but still we use
approximate non-Markovian QSD to simulate the Uhrig control dynamics \cite{KaoMou,YuStrunz2,JJing1}.

One of advantages of using non-Markovian QSD is its versatility in solving large environments with arbitrary finite memory times.  As an illustrative example,
the  explicit control dynamics measured by time-dependent
angular momentum and fidelity in the non-Markovian regime is solved by using the modified non-Markovian QSD equations.
Several scenarios for reinforcing the effectiveness of the regulation and control of three-level systems are considered such as  increasing the number of control pulses or engineering the environment to modify the environment memory time scale. Our method also allows an interesting extension to non-perturbative dynamical decoupled
in a non-Markovian regime.

In summary, with the versatility and the capability of the non-Markovian QSD equation,
the control processes can be substituted into the system dynamics consistently and be studied in detail.  More general extensions to multi-state atomic systems
are of importance where the non-Markovian QSD is known to be more powerful numerically.  It is also feasible to
investigate the non-ideal control pulse cases by using modified non-Markovian QSD equations \cite{Jing,Wang}.


\section* {Acknowledgements}
We thank  Prof.  J. H. Eberly,  Prof. Jiangbin Gong and Prof.  B. L. Hu for their useful discussions and the grant support
from the NSF PHY-0925174 and the AFOSR No. FA9550-12-1-0001, NSFC No. 11175110, Ikerbasque Foundation Startup,
 the Basque Government (Grant No. IT472-10), and the Spanish MEC (Project No. FIS2009-12773-C02-02).

\appendix

\section {Toggling frame of the dissipative three-level model under UDD control}

In this Appendix, we show the derivation of the interaction Hamiltonian in the Toggling frame. To start with, we list the  total Hamiltonian of the single three-level
model with two layers of UDD control pulse sequences according to Eq.~(\ref{totdissi}) and Eq.~(\ref{controldissi}) in Sec. III.
\begin{equation}
H_{\rm tot}=H_{\rm s}^{}+H_{\rm c1}+H_{\rm c2}+H_{\rm b}+H_{\rm i},
\end{equation}
where
\begin{align}
H_{\rm s}^{}  &  =\omega J_{z}, \non\\
H_{\rm c1} & =\sum_{j=1}^{N_1}\frac{\pi}{2}\de(t-T_j)P, \non\\
H_{\rm c2} & =\sum_{j=1}^{N_1}\sum_{k=1}^{N_2}\frac{\pi}{2}\de(t-T_{j,k})Q, \nonumber\\
H_{\rm e}  &  =\sum_{n}\omega_{n}b_{n}^{\dagger}b_{n}, \nonumber\\
H_{\rm i}  &  =\sum_{n}(g_{n}J_{-}b_{n}^{\dagger}+g_{n}^*J_{+}b_{n}) \non\\
& = \sum_n(g_nb_n^\da+g_n^*b_n)J_x+i(g_nb_n^\da-g_n^*b_n)J_y \non\\
&= B_xJ_x+B_yJ_y.
\end{align}

Rotating the reference frame with respect to $H_{\rm c1}$, the total Hamiltonian in the interaction picture is given by
 \beq H^{(1)}_{\rm tot} = e^{-i\int_0^TdsH_{\rm c1}}(H_{\rm s}+H_{\rm c2}+H_{\rm e}+H_{\rm i})e^{i\int_0^TdsH_{\rm c1}}. \eeq

By using the commutation relations,
\beqa
&[P,Q]=[P,J_x]=0, \non\\
&[P,[P,J_y]]=4J_y, \non\\
&[P,[P,J_z]]=4J_z,
\eeqa
we have
 \beqa
 e^{-i\int_{0}^{T}dsH_{c1}}Qe^{i\int_{0}^{T}dsH_{c1}} &=Q, \non\\
e^{-i\int_{0}^{T}dsH_{c1}}J_{x}e^{i\int_{0}^{T}dsH_{c1}} &=J_{x}, \non\\
e^{-i\int_{0}^{T}dsH_{c1}}J_{y}e^{i\int_{0}^{T}dsH_{c1}} &=p(t)J_{y}, \non\\
e^{-i\int_{0}^{T}dsH_{c1}}J_{z}e^{i\int_{0}^{T}dsH_{c1}} &=p(t)J_{z},
\eeqa
where the piecewise function $p(t)=\pm1$ with the value changes at time points $T_{j}$. Therefore the total Hamiltonian in the new frame becomes
\beqa
H^{\rm (1)}_{\rm tot} &=& H_{\rm s}^{(1)}+H_{\rm e}+H_{\rm c2}+H_{\rm i}^{(1)}, \non\\
H_{\rm s}^{\rm (1)} &=& f(t)\om J_z, \non\\
H_{\rm i}^{\rm (1)} &=& B_xJ_x+f(t)B_yJ_y.
\eeqa

Next we rotate the reference frame again with respect to $H_{\rm c2}$, the total Hamiltonian in this new interaction picture is
\beq H^{(2)}_{\rm tot} = e^{-i\int_0^TdsH_{\rm c2}}(H_{\rm s}^{(1)}+H_{\rm e}+H_{\rm i}^{(1)})e^{i\int_0^TdsH_{\rm c2}}.
\eeq

Now the commutation relations are
 \beqa
 &[Q,J_z]  =0, \non\\
 &[Q,[Q,J_x]] = 4J_x, \non\\
 &[Q,[Q,J_y]] = 4J_y,
 \eeqa
so we can get
 \beqa
 e^{-i\int_{0}^{T}dsH_{c2}}J_{y}e^{i\int_{0}^{T}dsH_{c2}} &=q(t)J_{x}, \non\\
e^{-i\int_{0}^{T}dsH_{c2}}J_{y}e^{i\int_{0}^{T}dsH_{c2}} &=q(t)J_{y}, \non\\
e^{-i\int_{0}^{T}dsH_{c2}}J_{z}e^{i\int_{0}^{T}dsH_{c2}} &=J_{z},
\eeqa
and  the $q(t)=\pm1$ with the value changes at time points $T_{j,k}$.  So using the rotating reference frame twice,  the total Hamiltonian becomes
 \beqa
H^{\rm (2)}_{\rm tot} &=& H_{\rm s}^{(1)}+H_{\rm e}+H_{\rm i}^{(2)} \non\\
H_{\rm i}^{\rm (2)} &=& q(t)B_xJ_x+q(t)p(t)B_yJ_y \non\\
&=& q(t)\left[\sum_n(g_nb_n^\da+g_n^*b_n)J_x+ip(t)(g_nb_n^\da-g_n^*b_n)J_y\right] \non\\
&=&  \sum_{n}(g_{n}b_{n}^{\dagger}L+g_{n}^*b_{n}L^\da)
\eeqa
 where the time-dependent Lindblad operator is  \beq L = q(t)\frac{1+p(t)}{2}J_-+q(t)\frac{1-p(t)}{2}J_+. \eeq

Finally, the effective Hamiltonian of the dissipative three-level model under two-layer UDD control sequences in the Toggling frame is
\beq  H^{\rm (2)}_{\rm tot}=p(t)\om J_z + \sum_{n}\omega_{n}b_{n}^{\dagger}b_{n}
 + \sum_{n}(g_{n}b_{n}^{\dagger}L+g_{n}^*b_{n}L^\da), \eeq
 This is the Eq.~(\ref{effective H dissi under control}) used in Sec.~III.

 \section {Dynamical equation for O operator}
 In the non-Markovian case, the linear stochastic Schr\"{o}dinger equation  was derived
in \cite{StrunzGisin}, it reads
 \begin{equation}
\frac{\partial}{\partial t}\psi_{t}=-iH_{\rm sys}\psi_{t}+Lz_{t}^{\ast}\psi
_{t}-L^{\dagger}
{\displaystyle\int\nolimits_{0}^{t}}
ds\alpha(t-s)\frac{\delta\psi_{t}}{\delta z_{s}^{\ast}}.
\end{equation}
It is noted that the above exact equation contains a time-nonlocal term. In order to find a time-local QSD equation, one can
introduce a time-dependent also noise-dependent operator $O(t,s,z^*)$, defined as
\beq
 \frac{\delta}{\delta z_{s}^{\ast}}\psi_{t}(z^*)=O(t,s,z^*)\psi_{t}(z^*),
\eeq
which can be determined from the consistency condition,
\beq
\frac{\delta}{\delta z_{s}^{\ast}}\frac{\partial \psi_{t}}{\partial t}=\frac{\partial}{\partial t}\frac{\delta \psi_{t}}{\delta z_{s}^{\ast}}.
\eeq
So we can derive the formal evolution equation for the operator $O(t,s,z^*)$,

\begin{equation} \label{cc}
\frac{\partial}{\partial t}O=[-iH_{sys}+Lz_{t}^{\ast}-L^{\dagger}\overline
{O},O]-L^{\dagger}\frac{\delta}{\delta z_{s}^{\ast}}\overline{O},
\end{equation}
where $\overline{O}(t,z^*)\equiv\int_0^tds\al(t,s)O(t,s,z^*)$. This equation
of motion for the O operator has to be solved with the
initial condition,
\beq
O(t,s=t,z^*)=L.
\eeq

For the three-level dephasing model under DD control described by Eq.~(\ref{effective H depashing under control}) in Sec. II, one can easily derive the exact $O$ operator
\beq
O(t,s,z^*)=f(t,s)J_z,
\eeq
with $f(t,s)=p(s)$. However, in Sec. III, for the second example with the three-level dissipative system and the control field, the explicit O operator cannot be determined.
In this case, we have to use a perturbative expansion in terms of noise $z^*$. This is a called weak noise perturbation, which means we choose the O operator containing  noise-free terms,
 \beqa
O(t,s)&=&f_1(t,s)J_-+f_2(t,s)J_+ \non\\
&& + f_3(t,s)J_zJ_-+f_4(t,s)J_zJ_+, \non\\
\overline{O}(t,s)&=& \int^t_0ds\al (t,s)O(t,s)\non\\
&=& F_1(t,s)J_-+F_2(t,s)J_+ \non\\
&&+ F_3(t,s)J_zJ_-+F_4(t,s)J_zJ_+,
\eeqa
where $F_i(i=1,2,3,4)=\int^t_0ds\al(t,s)f_i$.   Of course, the QSD equation still contains noise explicitly (more details are given by reference \cite{YuStrunz1}). By substituting the above two equations into Eq. (\ref{cc}), we can derive  a set of  differential equations for the coefficients as:
\beqa
 \frac{\partial}{\partial t}f_1 &=& (ip\om+l_2F_2-l_1F_3+l_2F_4)f_1-l_2F_1f_2,  \non\\
 \frac{\partial}{\partial t}f_2 &=& (-ip\om+l_1F_1-l_1F_3+l_2F_4)f_2-l_1F_2f_1, \non\\
 \frac{\partial}{\partial t}f_3 &=& (ip\om-l_1F_1+l_1F_3-l_2F_4)f_3 \non\\
 &&+(l_1F_1-2l_1F_3+l_2F_2+2l_2F_4)f_1 \non\\
 &&-2l_2F_1f_2-l_2F_1f_4, \non\\
 \frac{\partial}{\partial t}f_4 &=& (-ip\om-l_2F_2+l_1F_3-l_2F_4)f_4 \non\\
 &&-(l_1F_1-2l_1F_3+l_2F_2+2l_2F_4)f_2 \non\\
 &&+2l_1F_2f_1 -l_1F_2f_3,
  \eeqa
with the initial conditions:
\beqa
f_1(t,s=t) &=& l_1(t), \non\\
f_2(t,s=t) &=& l_2(t), \non\\
f_3(t,s=t) &=& 0, \non\\
f_4(t,s=t) &=& 0.
\eeqa

In this paper, we choose the Ornstein-Ulenback type of correlation function $\al(t,s)=\frac{\ga}{2}e^{-\ga|t-s|}$, so the equations of motion for $F_i$ are:
\beqa
 \frac{\partial}{\partial t}F_1 &=&\frac{\ga}{2}l_1+(ip\om-l_1F_3+l_2F_4-\ga)F_1, \non\\
 \frac{\partial}{\partial t}F_2 &=&\frac{\ga}{2}l_2+(-ip\om-l_1F_3+l_2F_4-\ga)F_2, \non\\
 \frac{\partial}{\partial t}F_3 &=&(ip\om-3l_1F_1+l_1F_3-l_2F_4-\ga)F_3 \non\\
 &&+(l_1F_1-l_2F_2+l_2F_4)F_1, \non\\
 \frac{\partial}{\partial t}F_4 &=& (-ip\om-3l_2F_2+l_1F_3-l_2F_4-\ga)F_4 \non\\
 &&+(l_1F_1-l_2F_2+l_1F_3)F_2,
  \eeqa
  with $F_i(0)=0$.  Consequently, the $O$ operator and the QSD equation are fully determined.

\end{document}